\documentclass[aps, prl,twocolumn,superscriptaddress,showpacs]{revtex4-1}

\usepackage{graphicx}
\usepackage{hyperref}
\usepackage{amsmath}
\usepackage{amssymb}
\usepackage{epstopdf}
\newcommand{\ket}[1]{\vert #1 \rangle}

\begin{document}

\title{Optical Synthesis of Large-Amplitude Squeezed \\Coherent-State Superpositions with Minimal Resources}

\author{K. Huang}
\affiliation{Laboratoire Kastler Brossel, UPMC-Sorbonne Universit\'es, CNRS, ENS-PSL Research University, Coll\`ege de France, 4 place
Jussieu, 75005 Paris, France}
\affiliation{State Key Laboratory of Precision Spectroscopy,
East China Normal University, Shanghai 200062, China}
\author{H. Le Jeannic}
\affiliation{Laboratoire Kastler Brossel, UPMC-Sorbonne Universit\'es, CNRS, ENS-PSL Research University, Coll\`ege de France, 4 place
Jussieu, 75005 Paris, France}
\author{J. Ruaudel}
\altaffiliation[]{Present address: Laboratoire Aim\'e Cotton, CNRS, Univ Paris-Sud, ENS Cachan, 91405 Orsay, France.}
\affiliation{Laboratoire Kastler Brossel, UPMC-Sorbonne Universit\'es, CNRS, ENS-PSL Research University, Coll\`ege de France, 4 place
Jussieu, 75005 Paris, France}
\author{V. B. Verma}
\affiliation{National Institute of Standards and Technology, 325 Broadway, Boulder, CO 80305, USA}
\author{M. D. Shaw}
\affiliation{Jet Propulsion Laboratory, California Institute of Technology,
4800 Oak Grove Dr., Pasadena, California 91109, USA}
\author{F. Marsili}
\affiliation{Jet Propulsion Laboratory, California Institute of Technology,
4800 Oak Grove Dr., Pasadena, California 91109, USA}
\author{S. W. Nam}
\affiliation{National Institute of Standards and Technology, 325 Broadway, Boulder, CO 80305, USA}
\author{E Wu}
\affiliation{State Key Laboratory of Precision Spectroscopy,
East China Normal University, Shanghai 200062, China}
\author{H. Zeng}
\affiliation{State Key Laboratory of Precision Spectroscopy,
East China Normal University, Shanghai 200062, China}
\author{Y.-C. Jeong}
\affiliation{Laboratoire Kastler Brossel, UPMC-Sorbonne Universit\'es, CNRS, ENS-PSL Research University, Coll\`ege de France, 4 place
Jussieu, 75005 Paris, France}
\author{R. Filip}
\affiliation{Department of Optics, Palack\'y University, 17. listopadu 1192/12, 77146 Olomouc, Czech Republic}
\author{O. Morin}
\altaffiliation[]{Present address: Max-Planck-Institut f\"ur Quantenoptik, Hans-Kopfermann-Str. 1, D-85748 Garching, Germany.}
\affiliation{Laboratoire Kastler Brossel, UPMC-Sorbonne Universit\'es, CNRS, ENS-PSL Research University, Coll\`ege de France, 4 place
Jussieu, 75005 Paris, France}
\author{J. Laurat}
\email{julien.laurat@upmc.fr}
\affiliation{Laboratoire Kastler Brossel, UPMC-Sorbonne Universit\'es, CNRS, ENS-PSL Research University, Coll\`ege de France, 4 place
Jussieu, 75005 Paris, France}

\date{\today}

\begin{abstract}
We propose and experimentally realize a novel versatile protocol that allows the quantum state engineering of heralded optical coherent-state superpositions. This scheme relies on a two-mode squeezed state, linear mixing and a $n$-photon detection. It is optimally using expensive non-Gaussian resources to build up only the key non-Gaussian part of the targeted state. In the experimental case of a two-photon detection based on high-efficiency superconducting nanowire single-photon detectors, the freely propagating state exhibits a $67\%$ fidelity with a squeezed even coherent-state superposition with a size $|\alpha|^2$=3. The demonstrated procedure and the achieved rate will facilitate the use of such superpositions in subsequent protocols, including fundamental tests and optical hybrid quantum information implementations.
\end{abstract}


\maketitle

The optical hybrid approach to quantum information, which consists in combining the traditionally separated discrete- and continuous-variable states and operations, have seen important developments in recent years \cite{vanLoock2011,Andersen2014}. First demonstrations of hybrid protocols offering novel capabilities have been achieved, including a long-distance single-photon entanglement witness based on local homodyne detections \cite{Morin2013,Morin2014} or the deterministic continuous-variable teleportation of discrete quantum bits \cite{Takeda2013}. Quantum repeater architectures have also been investigated \cite{Sangouard2010,Brask2010} and the recent demonstrations of entanglement between particle-like and wave-like qubits \cite{Jeong2014,MorinNP} open the promise of heterogeneous networks based on the combination of both encodings. This hybridization also finds a variety of unique applications in linear optical quantum computing \cite{Lund2008,Lee2013} and quantum metrology \cite{Joo2011}.

Within this context, a considerable effort has been dedicated to the generation of highly non-Gaussian states of light \cite{Andersen2014}. Specifically, free-propagating coherent-state superpositions (CSS), also referred as optical Schr\"odinger cat states, are an essential resource. Such states of the form $\ket{\alpha}\pm\ket{-\alpha}$ consisting in a superposition of two coherent states with opposite phases and mean photon number $|\alpha|^2$, plays the role of qubits in the coherent state basis \cite{Jeong2002,Ralph2003,Ralph2010}. The size $|\alpha|^2$ of the superposition is a critical parameter as directly related to the overlap between the two coherent-state components. A value $|\alpha|^2=2$ gives already an overlap $|\langle\alpha|-\alpha\rangle|^2=e^{-4|\alpha|^2}\simeq 3.10^{-4}$. Beyond their fundamental significance for hybrid Bell tests and the study of mascropicity for quantum optical states \cite{AndersenMicro,Kwon2015}, the generation of CSS with this minimal size and a generation rate large enough to allow subsequent operations will open a wealth of possible protocols and gate implementations \cite{Marek,vanLoock2011,Andersen2014}. However, such generation remains very challenging.

In this endeavor, various optical circuits have been developed to generate CSS using non-Gaussian resources. The first seminal scheme consisted in subtracting a single-photon from a single-mode squeezed vacuum \cite{Ourjoumtsev2006,Nielsen2006,Wakui2007}. This operation can be performed by tapping a part of squeezed light and detect it with a single-photon detector. This process results in heralding a squeezed single-photon, which exhibits a high fidelity with an odd CSS with $|\alpha|^2\sim1$. In order to reach larger $\alpha$ values, other protocols have then been implemented.  Two-photon subtraction operated on squeezed light, with \cite{Takahashi2008} or without \cite{Gerrits} time separation, has led to values $|\alpha|^2$ close to 2. An alternative method has provided a 3 dB-squeezed CSS with $|\alpha|^2=2.6$. For this purpose, a two-photon Fock state is first heralded, split on a beamsplitter and one output is measured by homodyne detection \cite{Ourjoumtsev2007}. A quadrature measurement within a given acceptance window is used as third conditioning. Recently, a similar approach based on the interference of two single-photons and homodyne conditioning has led to the same result \cite{Etesse2014}. In all these experiments, the generation rate is low, at the Hz level or below, precluding their use in subsequent protocols. This limited rate comes from either operation on single-mode squeezed light, where intrinsically only a small part can be tapped, or the requirement of three cascaded conditioning procedures.

In this Letter, we propose and implement a scheme enabling the heralded generation of coherent-state superpositions with the minimal required resources. This state engineering relies on two-mode squeezed vacuum and a $n$-photon detection performed on one of the modes. No coherent displacement is required, in contrast to previous works  \cite{Bimbard2010,Yukawa2013}.  The process allows to optimize the formation of the CSS in a versatile way, focusing all the non-Gaussian resources to prepare only the non-Gaussian part of the state. Experimentally, using a type-II optical parametric oscillator (OPO), we generate a state showing a 67\% fidelity with a squeezed even CSS with a size $|\alpha|^2$=3. The preparation rate is two orders of magnitude larger than achieved heretofore with the aforementioned schemes. We also provide detailed characterization of the process enabling to change the size of the superposition and its squeezing. These observations are made possible by the combination of a large escape efficiency OPO and high-efficiency superconducting single-photon detectors.

\begin{figure}[t!]
\includegraphics[width=0.93\columnwidth]{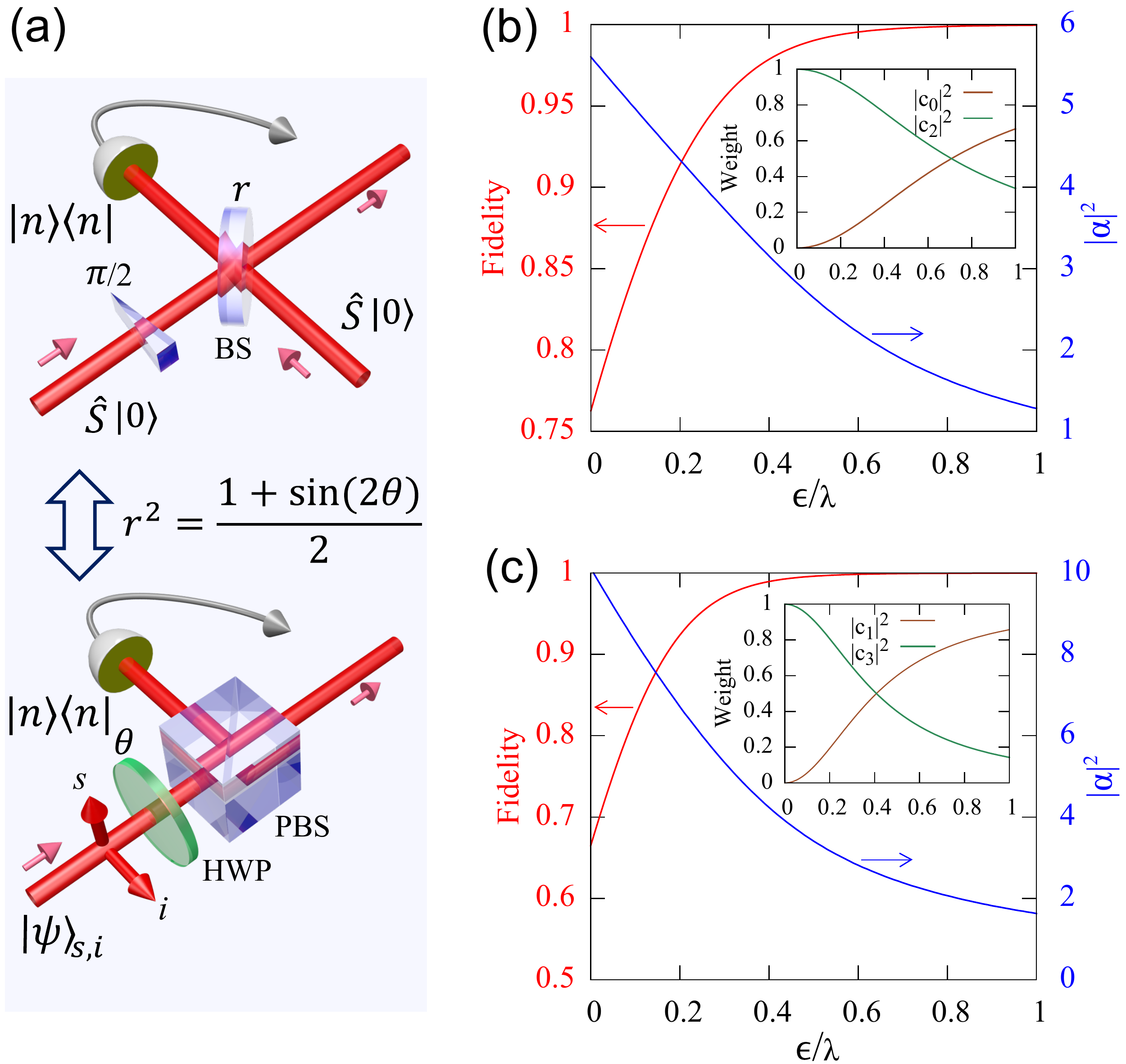}
\caption{(color online). Schemes for generating $\alpha\ket{n-2}~+~\beta\ket{n}$ and theoretical simulations. (a) Two squeezed vacua are mixed on a tunable beamsplitter. The detection of $n$ photons in one of the outputs heralds the generation. Equivalently, the scheme can be implemented directly from a two-mode squeezed vacuum. The two orthogonally-polarized modes are separated with a tunable mixing $\epsilon$ induced by a rotated half-wave plate and a polarizing beamsplitter, with $\epsilon=\sin(2\theta)$. The angle $\theta=0^{\circ}$ corresponds to the perfect separation, leading to the generation of a $n$-photon state. (b) Two-photon detection: fidelity with a squeezed even CSS as a function of the ratio $\epsilon/\lambda$. The fidelity is optimized by the size $|\alpha|^2$ and the squeezing of the target state. The inset gives the weights of the vacuum and two-photon components. (c) Three-photon detection and fidelity with a squeezed odd CSS.}
\label{fig1}
\end{figure}

We first describe the proposed scheme, as illustrated on Fig. \ref{fig1}(a). Two squeezed vacua, $\hat{S}(\xi)\ket{0}$, with a squeezing factor $s=e^{-2\xi}$ and $\lambda=\tanh(\xi)$, are first overlapped with a $\pi/2$ phase-shift on a tunable beamsplitter. This configuration has been widely used in the balanced case as it provides a two-mode squeezed vacuum state for subsequent operations, such as continuous-variable teleportation \cite{Furusawa98,Bowen03}. In contrast, the beamsplitter here is asymmetric and a photon-counting measurement is performed on one of the outputs to herald the generation of the synthesized superposition. In the following, we denote $\epsilon\ll 1$ the beamsplitter asymmetry, i.e. $r=\sqrt{(1+\epsilon)/2}$. In the ideal case of a photon-number resolving detector, the detection of $n$ photons in the conditioning channel heralds the state
\begin{eqnarray}\label{core}
\ket{\Psi}=\frac{1}{\sqrt{n(n-1)\epsilon^2+\lambda^2}}\left( \epsilon\sqrt{n(n-1)}\ket{n-2}+\lambda\ket{n}\right).\nonumber
\end{eqnarray}
This expression stays valid with multiplexed on-off single-photon detectors in the limit of low squeezing in order to neglect the multiple-photon components.

Equivalently this scheme can be realized by starting from a two-mode squeezed state, i.e. photon-number correlated signal and idler beams $\ket{\Psi}_{s,i}~=~\sqrt{1-\lambda^2} \sum \lambda^n\ket{n}_s\ket{n}_i$, as generated for instance by a type-II OPO below threshold. In this case, the modes are orthogonally polarized and the mixing is realized by separating the beams after a small polarization rotation induced by a half-wave plate. The mixing parameter is therefore given by $\epsilon=\sin(2\theta)$. The relative phase between the two CSS components can be easily controlled by birefringent elements inserted before the mode splitting.

Numerical simulations are displayed in Figs. \ref{fig1}(b) and (c) for a two- and a three-photon heralding detection. The fidelity is calculated with an ideal squeezed, odd or even, CSS as given by:
\begin{eqnarray}\label{CSS}
\hat{S}(\xi')\ket{\textrm{CSS}_{\pm}}=\frac{1}{\sqrt{2(1\pm e^{-2\alpha^2})}}\hat{S}(\xi')\left(\ket{\alpha}\pm\ket{-\alpha}\right).\nonumber
\end{eqnarray}
For each point, the fidelity is optimized by the size and the squeezing factor $s'=e^{-2\xi'}$. Size $|\alpha|^2$ as large as 3 for $n=2$ and 5 for $n=3$ can be obtained with fidelities above 98\%. 

This procedure is an example of a very general method to optimize the preparation of non-Gaussian states with Gaussian operations \cite{Radim2009}. This study proved that Gaussian operations, such as squeezing, can reduce the required non-Gaussian resources for state preparation. By identifying the essential non-Gaussian operational cost, they defined a set of core non-Gaussian states that are good approximations of the targeted state via only a squeezing operation. This strategy can lead to better fidelities with fewer photon detections than methods based only on state truncation. Within this approach, the work reported here provides the best approximation of such core states for the targeted CSS given available $n$-photon detection. In other words, it enables to produce the core state that requires the minimum number of photon detections to be prepared. The remaining squeezing operation can be implemented using for instance the universal squeezer already tested for non-Gaussian states \cite{Furusawa2014}. Moreover, the core states are generally more robust against a damping than the superposition of coherent states itself \cite{Radim2013}. They are therefore more suitable for the transmission and storage of non-Gaussian states.

\begin{figure}[t!]
\includegraphics[width=0.93\columnwidth]{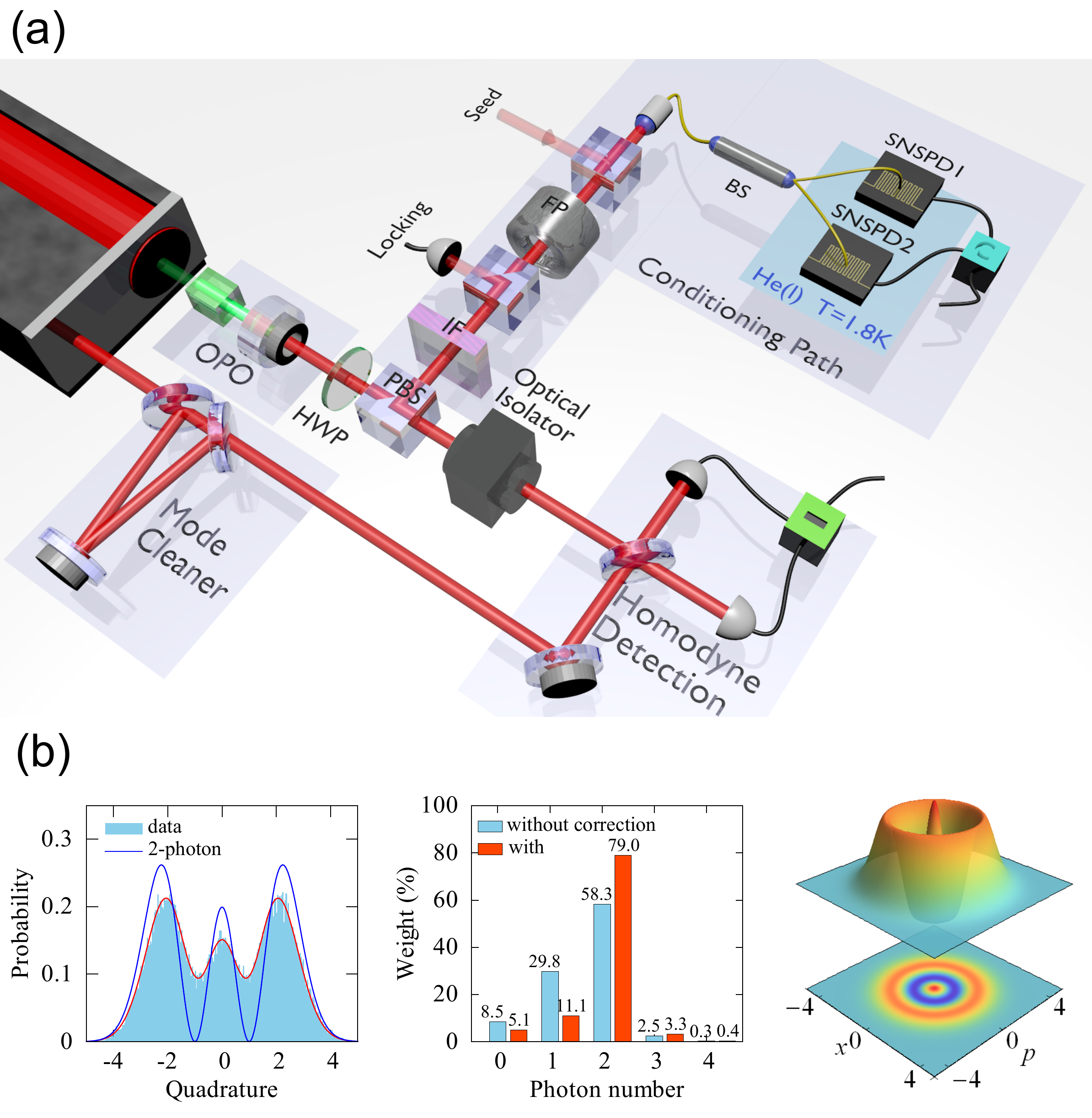}
\caption{(color online). Experimental setup. (a) A two-mode squeezed vacuum generated by a type-II OPO is split on a tunable polarizing beamsplitter. One of the outputs, which is frequency filtered by an interferential filter (IF) and a cavity (FP), impinges on a fiber beamsplitter. Coincidence events from high-efficiency superconducting single-photon detectors herald the preparation. The resulting state is characterized by homodyne detection, with an overall efficiency $\eta=85\%$. (b) Two-photon state generation when signal and idler are separated without mixing ($\theta=0^{\circ}$). The first plot gives the measured quadrature distribution while the second one provides the diagonal elements of the density matrix, with and without correction for detection losses. The right figure displays the corresponding Wigner function without correction, with a negative ring value $W=-0.13\pm 0.01$. The heralding rate reaches 200 Hz.}
\label{fig2}
\end{figure}

We now turn to the experimental realization. The setup is sketched on Fig. \ref{fig2}(a). The two-mode squeezed vacuum is generated by a type-II OPO pumped far below threshold by a continuous-wave, frequency-doubled Nd:YAG laser (Diabolo Innolight) \cite{MorinOL}. The OPO is made of a triply-resonant semi-monolithic linear cavity where the input mirror is directly coated on one face of a 10-mm KTP crystal (Raicol) and the output coupler is a 38-mm-curved mirror. The input mirror exhibits an intensity reflection of 95\% for the 532 nm pump and high reflection at 1064 nm, while the output coupler is highly reflective for the pump and has a 90\% reflection for the infrared, leading to a bandwidth equal to 60 MHz and an escape efficiency estimated to $\eta_{OPO}\sim0.9$.

At the OPO output, the orthogonally polarized signal and idler impinge on a half-wave plate and a polarizing beamsplitter, which enable to mix the two modes and separate the resulting orthogonal polarizations. An angle $\theta=0^{\circ}$ corresponds to separate signal and idler without coupling them. The reflected mode, which is used for the heralding, is frequency-filtered to remove the non-degenerate modes due to the OPO cavity. This filtering is based on a 0.5-nm interferential filter and a 320-MHz-bandwidth Fabry-Perot cavity \cite{MorinOL,MorinJove}. The filtered light is then split on a 50/50 fiber beamsplitter and detected by two superconducting nanowire single-photon detectors (SNSPDs) based on tungsten silicide (WSi) \cite{Marsili2013}. The optical stack of the SNSPDs was designed for maximum absorption at 1064 nm by optimizing the thickness of the dielectric layers. Details of the detectors system will be reported elsewhere. The system detection efficiency reaches 85\% while dark count rate is below 10 cps, an important feature for high-fidelity state generation \cite{DAuria2011,DAuria2012}.

\begin{figure*}[t!]
\includegraphics[width=1.9\columnwidth]{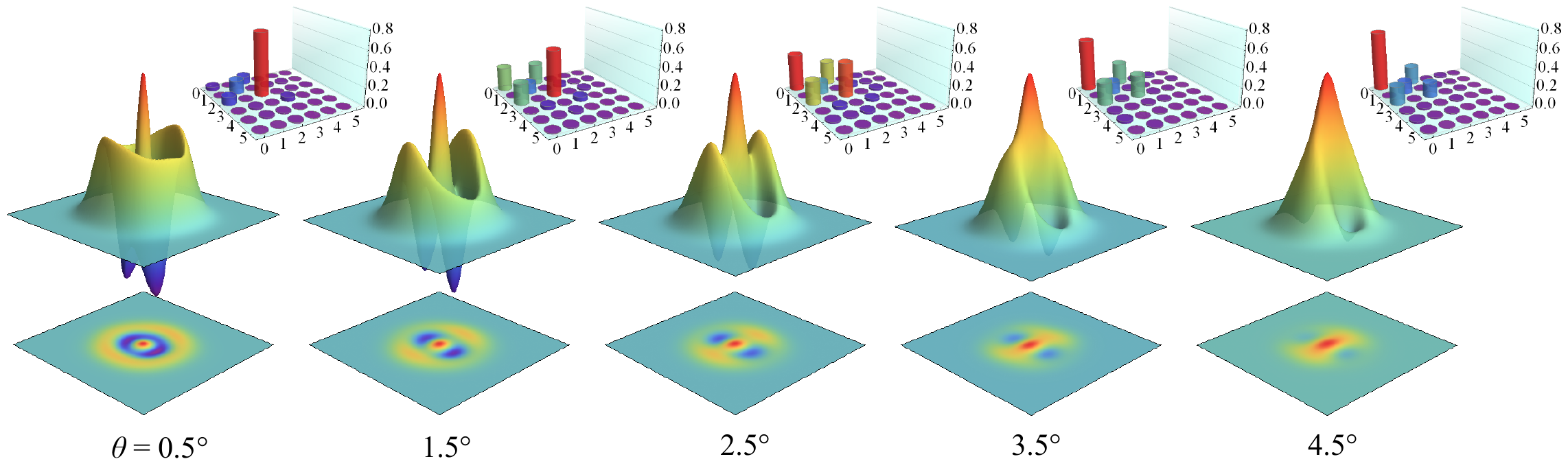}
\caption{(color online). Experimental results. The plots give the density matrices (absolute values) and the associated Wigner functions for incremental values of the half-wave plate angle $\theta$ ($\epsilon$ between 0.017 and 0.16). The results are corrected for the $\eta=85\%$ detection efficiency. For $\theta=1.5^{\circ}$, the negative peaks of the Wigner function reach $W=-0.27\pm0.01$.}
\label{fig3}
\end{figure*}

As the experiment is performed in the continuous-wave regime, detection events can occur at different times. Here, the accepted coincidence window between the two heralding triggers is set to 0.8 ns, much smaller than the typical time given by the inverse of the OPO bandwidth. As a result, the temporal mode in which the state is generated is given by a double-decaying exponential profile with a width at $1/e$ equal to 30 ns \cite{MorinMode}. The heralded state is finally characterized by quantum state tomography via homodyne detection. Quadrature values from 50000 measurements are processed with a maximum likelihood algorithm \cite{LvovskyTomo}, which provides the state density matrix and the corresponding Wigner function.

The experiment is carried out in a cyclic fashion with 50-ms data acquisition and 50-ms locking and calibration periods. During the locking time, two auxiliary beams, otherwise blocked by mechanical shutters, are used. A first counter-propagating beam enables to lock the filtering cavity using a microcontroller-based technique \cite{HuangLocking}. A second weak beam is injected into the OPO cavity. This beam is amplified or de-amplified depending on the relative phase with the pump. To set this phase, the intensity is locked at a constant level. The interference between the seed and the local oscillator is then used to access the phase of the quadrature measured by the homodyne detection.

As a first experimental characterization of the system, we consider the case $\theta=0^{\circ}$, which should lead to the generation of a two-photon Fock state. Figure \ref{fig2}(b) gives the measured quadrature distribution, the diagonal elements of the reconstructed density matrix and the corresponding Wigner function. Without correction for losses, the two-photon component reaches a value of 58\%. This value is the highest fidelity reported to date. As the OPO is pumped far below threshold, the three-photon component is kept below 3\%.  By correcting for detection losses, the two-photon component is 79 \%, as mainly limited by the square of the OPO escape efficiency $\eta_{OPO}^2\sim0.8$.

We now investigate the evolution of the heralded state when the signal and idler mixing is increased. Figure \ref{fig3} shows the density matrices and the corresponding Wigner functions for different values of the waveplate angle $\theta$, from $0.5^{\circ}$ to $4.5^{\circ}$. As can be seen, the mixing strongly modifies the generated states, which are not anymore phase invariant and exhibit two negative peaks in the Wigner functions. These peaks correspond to oscillations in phase space between the two coherent-state components. Remarkably, due to the efficient SNSPDs and the strategy based on two-mode squeezed light, the heralding rate reaches 200 Hz. This rate is at least 2 orders of magnitude larger than the ones achieved heretofore in experiments usually based on photon subtraction operated on squeezed light or involving three conditioning steps.

\begin{figure}[b!]
\includegraphics[width=0.92\columnwidth]{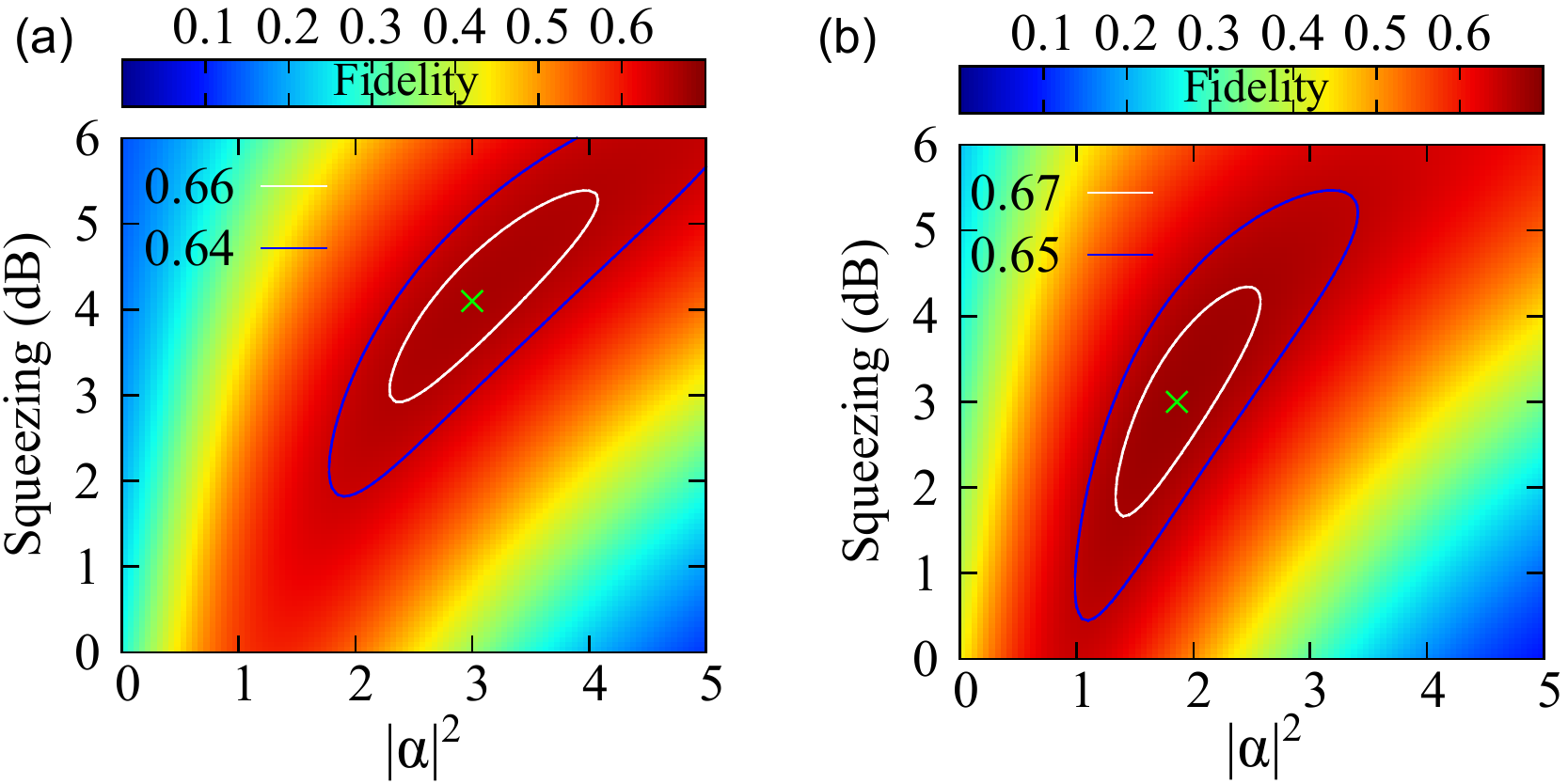}
\caption{(color online). Fidelity between the generated state and a squeezed even CSS, for (a) $\theta=1.5^{\circ}$ and (b) $\theta=2.5^{\circ}$. The plots give the calculated fidelity as a function of $|\alpha|^2$ and the squeezing in dB. The green crosses indicate the maximal fidelities. For $\theta=1.5^{\circ}$, the fidelity reaches 0.67 with an even CSS with a size $|\alpha|^2=3$ and a 4-dB squeezing. For $\theta=2.5^{\circ}$, the fidelity reaches 0.68 with an even CSS with a size $|\alpha|^2=1.9$ and a 3-dB squeezing.}
\label{fig4}
\end{figure}

Turning to a characterization of the generated state, we present in Fig. \ref{fig4} its fidelity with an ideal squeezed even CSS, for two different angles. The fidelity is displayed as a function of the size $|\alpha|^2$ and its squeezing in dB. For $\theta=1.5^{\circ}$, shown in Fig. \ref{fig4}(a), a fidelity $F=0.67\pm 0.01$ is obtained for $|\alpha|^2=3$ and a squeezing of 4 dB. The generated state thereby exhibits the highest amplitude and fidelity reported to date for free-propagating CSS. Furthermore, the scheme is versatile. Indeed, by simply adjusting the mixing ratio $\epsilon$, i.e. the wave-plate angle, the properties of the state can be engineered. For instance, as seen in Fig. \ref{fig4}(b), the state generated with $\theta=2.5^{\circ}$ exhibits a maximal fidelity $F=0.68\pm0.01$ for $|\alpha|^2=1.9$ and a squeezing of 3 dB. For any targeted squeezing, one can find the optimal angle to be used. 

We finally note that, given the escape efficiency of the OPO, the corrected fidelity is expected to reach a value close to 80\%, similarly to the one obtained for the two-photon state displayed in Fig. \ref{fig2}(b). The discrepancy mainly comes from the strong phase sensitivity of such superposed states. Due to phase calibration performed only every few seconds at the beginning of each acquisition sequence and transient perturbation of the experiment by mechanical shutters, errors on the phase calibration reduce the achieved fidelity.

In conclusion, we have proposed and implemented a scheme enabling a versatile quantum state engineering of squeezed coherent-state superpositions, which are essential highly non-classical states in the emerging field of optical hybrid quantum information processing. Given a $n$-photon detection, the protocol provides the best approximation of the non-Gaussian core state for the targeted CSS. Experimentally, in the case of a two-photon heralding, a 4 dB-squeezed even CSS state is obtained with $|\alpha|^2=3$ in a well-controlled spatio-temporal mode. Importantly, this scheme is based on twin beams, and not on single-mode squeezed light or cascaded conditionings that usually strongly limit the heralding rate. The achieved preparation rate makes these states immediately suitable for subsequent experiments where they can be used as initial resources, such as in test implementations of all-optical non-Gaussian gates for quantum communication networks \cite{Jeong2002,Ralph2003,Marek}. From this perspective, it is also advantageous that the generated squeezed CSS can be optimally adapted for transmission in optical channels and storage in atomic memories \cite{Radim2013}. Further improvements of the non-Gaussian core states can be achieved by feasible extensions with developing photon-number resolving detectors. Beyond direct applications, the generated CSS and its extension to larger core states can also be used to investigate processes connected to fundamental quantum-classical transition \cite{Zurek}.
\\

\begin{acknowledgments}
This work was supported by the ERA-Net CHIST-ERA (QScale) and the European Research Council (Starting Grant HybridNet). Part of this research was carried out at the Jet Propulsion Laboratory, California Institute of Technology, under a contract with the National Aeronautics and Space Administration. V.B.V. and S.W.N. acknowledge partial funding for detector development from the DARPA Information in a Photon (InPho) and QUINESS programs. K.H. was supported by the Foundation for the Author of National Excellent Doctoral Dissertation of China (PY2012004) and the China Scholarship Council, and R.F. by the grant GA14-36681G of the Czech Science Foundation. The authors also acknowledge the technical assistance provided by B.~Huard's Quantum Electronics group at LPA, Paris. J.L. is a member of the Institut Universitaire de France.\\
\end{acknowledgments}

\end{document}